\documentclass{kluwer}

\usepackage{psfig}

\begin{document}

\begin{article}
\begin{opening}         
\title{Dust and Nebular Emission in
Star Forming Galaxies}
\author{Pasquale \surname{Panuzzo}\email{panuzzo@sissa.it}}  
\institute{S.I.S.S.A., Via Beirut 4, I-34014 Trieste, Italy}
\author{Alessandro \surname{Bressan}}
\author{Gian Luigi \surname{Granato}}
\institute{Osservatorio Astronomico, Vicolo
dell'Osservatorio 5, I-35122 Padova, Italy}
\author{Laura \surname{Silva}}
\institute{Osservatorio Astronomico, Via G. B. Tiepolo
11, I-34131 Trieste, Italy}
\author{Luigi \surname{Danese}}
\institute{S.I.S.S.A., Via Beirut 4, I-34014 Trieste, Italy}

\begin{abstract}
Star forming galaxies exhibit a variety of physical
conditions, from quiescent normal spirals to the 
most powerful dusty starbursts.
In order to study these complex systems, we need a suitable tool
to analyze the information coming from observations
at all wavelengths. We present a new spectro-photometric model 
which considers in a consistent way starlight as
reprocessed by gas and dust. We discuss
preliminary results to interpret some observed properties
of VLIRGs.
\end{abstract}

\end{opening}

%%%%%%%%%%%%%%%%%%%%%%%%%%%%%%%%%%%%% 2

\section{Introduction: why we care about nebular emission}
A common property of star forming galaxies
is the presence of emission lines in the spectrum.
Emission lines are produced by gas that is ionized by UV
radiation emitted from young massive stars.
Star-forming regions are the sites where the nebular emission is
produced, but these regions are also dusty environments.
It seems a natural goal for a model of star-forming galaxies
to treat contextually the nebular emission 
and dust processing.

We present the implementation of nebular emission
calculation in the spectrophotometric code GRASIL 
\cite{silva}. GRASIL is a code built to simulate
the SED of galaxies including a careful treatment of
starlight dust
reprocessing.

The study of emission lines with GRASIL 
will give us many further constraints
about:
\begin{itemize}
\item
The dust obscuration:
the optical thickness of star-forming regions can be derived 
from line ratio while it is very difficult to constrain optical thickness
of ISM only from continuum.
\item 
The star formation rate:
emission lines are believed to be powerful SFR estimators. We can now compare
SFR derived from nebular lines with the results from FIR, UV etc.
\end{itemize}

%%%%%%%%%%%%%%%%%%%%%%%%%%%%%%%%%%%%% 3

\section{The model}
%\subsection{Dust and stars}
In our model, the galaxy is composed by two
main components: the bulge and the disk. 
Dust within the disk is divided in:
i) molecular clouds (MCs), where the star formation is active,
ii) diffuse medium (or cirrus).

Young stars are supposed to be born into the MCs,
and leave them progressively as their age increases.
This is modelled by considering that a suitable fraction 
of the light of SSP is radiated inside the MCs.
This fraction is a function of age, parametrized by an
escaping time. 

So we have two populations: A) young stars inside MCs,
B) older stars outside MCs.

A molecular cloud is modelled as a thick spherical shell
of dense gas (and dust) around a central point source representing all the
star content of the cloud.
The MCs, as well as old stars, are embedded in the diffuse medium.
For more details on energy transport, refer to Silva et
al. \cite*{silva} or Silva et
al., this conference.

%%%%%%%%%%%%%%%%%%%%%%%%%%%%%%%%%%%%% 5

%\subsection{Computing Nebular Emission}
To calculate emission lines
we used the photoionization code CLOUDY 94 \cite{fer96}.

We built a library of HII regions in order to avoid the computation 
of entire nebular emission via CLOUDY at every run of the code GRASIL.

The HII regions of the library are calculated for
different density, metallicity and radii.
The SEDs of ionizing sources of HII regions are obtained by integrating the 
light of each of the two populations for several assumption of
escaping time and IMF. 

The SEDs of ionizing sources are parametrized
by the number of ionizing photons for HI, HeI and OII (Q(H), Q(He), and Q(O)).
Q(H) gives the mass of ionized gas, while the ratios Q(He)/Q(H) and Q(O)/Q(H)
are related to the hardness of ionizing flux and to the degree of ionization.
In table \ref{elenco} it is reported the list of computed lines.

GRASIL calculates the Qs of 
each population (inside and outside the MC), interpolates on the library
of HII regions, and computes the nebular emission for
each population.
Then, the nebular emission is extincted  in the same way as the population
that produced it.

\begin{table}
\caption[]{Computed lines: Hydrogen recombination lines
(upper pannel), Elium and metal lines (lower pannel)}
\label{elenco}
\tiny
\begin{tabular}{llllll}
\hline
Ly$\alpha$ 1216&Ly$\beta$ 1025&Ly$\gamma$ 972&Ly$\delta$ 949&Ly 937&Ly 930\\
Ly 926&Ly 922&H$\alpha$6563 &H$\beta$ 4861&H$\gamma$ 4340&H$\delta$ 4102\\
H 3970&H 3889&H 3835&H 3798&Pa$\alpha$ 18752&Pa$\beta$ 12819\\
Pa$\gamma$ 10939&Pa$\delta$ 10050&Pa 9546&Pa 9229&Pa 9015&Pa 8863\\
Br$\alpha$ 40515&Br$\beta$  26254&Br$\gamma$ 21657&Br$\delta$ 19447&Br 18175&Br 17363\\
Br 16808&Br 16408&Pf$\alpha$ 74585&Pf$\beta$  46529&Pf$\gamma$  37398&Pf$\delta$  32964\\
Pf 30386&Pf 28724&Pf 27577&Pf 26746&Hu$\alpha$ 123690&Hu$\beta$ 75011\\
Hu$\gamma$ 59071&Hu$\delta$  51277&Hu 46716&Hu 43756&Hu 41700&Hu    40201\\
\hline
HeII 1640&HeII 1217&HeII 1085&HeII 4686&HeII 3203&HeII 2733\\
HeII 2511&HeI  4471&HeI  5876&HeI  6678&HeI   10830&HeI    3889\\
HeI  7065&$[$CI$]$9850&$[$CI$]$8727&$[$CI$]$4621&$[$CI$]$609$\mu$m&$[$CI$]$369$\mu$m\\
$[$CII$]$157.7$\mu$m &CII$]$2326 &CIII$]$1908&$[$NI$]$5199   &$[$NI$]$3466   &$[$NI$]$10400\\
$[$NII$]$6584&$[$NII$]$6548&$[$NII$]$5755&$[$NII$]$122$\mu$m&$[$NII$]$205$\mu$m&NII$]$2141\\
$[$NIII$]$57$\mu$m&$[$OI$]$6300&$[$OI$]$6363&$[$OI$]$5577&$[$OI$]$63$\mu$m&$[$OI$]$145$\mu$m\\
$[$OII$]$3727  &$[$OII$]$7325  &$[$OII$]$2471&OIII$]$1663  &$[$OIII$]$5007 &$[$OIII$]$4959\\
$[$OIII$]$4363 &$[$OIII$]$2321&$[$OIII$]$88$\mu$m&$[$OIII$]$52$\mu$m&$[$OIV$]$26$\mu$m&$[$NeII$]$13$\mu$m\\
$[$NeIII$]$15.5$\mu$m&$[$NeIII$]$36$\mu$m &$[$NeIII$]$3869&$[$NeIII$]$3967&$[$NeIII$]$3343&$[$NeIII$]$1815\\
$[$NeIV$]$2424&$[$NeIV$]$4720 &MgII2800   &$[$SiII$]$35$\mu$m&$[$SII$]$10330 &$[$SII$]$6731\\
$[$SII$]$6717  &$[$SII$]$4070&$[$SII$]$4078  &$[$SIII$]$19   &$[$SIII$]$33.5 &$[$SIII$]$9532 \\ 
$[$SIII$]$9069 &$[$SIII$]$6312 &$[$SIII$]$3722 &$[$SIV$]$10.4$\mu$m&$[$ArII$]$69850&$[$ArIII$]$7135\\
$[$ArIII$]$7751&$[$ArIII$]$5192&$[$ArIII$]$3109&$[$ArIII$]$3005&$[$ArIII$]$22$\mu$m&$[$ArIII$]$9$\mu$m\\
\hline
\end{tabular}
\end{table}

\section{An application to Very Luminous InfraRed Galaxies}
VLIRGs ($L_{\rm IR} \geq 10^{11.5}L_\odot$)
are the most powerful star forming galaxies at low redshift.
Recent results show that: VLIRGs do not follow the Meurer's relation
$L_{\rm FIR}/L_{\rm FUV}$ versus UV spectral index $\beta$
\cite{meu99,meu00}; SFR computed from H$\alpha$, even
corrected for extinction, is always lower than SFR derived from FIR
\cite{pog00}.
%3) The ratio  $L_{{\rm H}\alpha_0}/L_{bol}$ in VLIRGs is significantly 
%lower than expected (Meurer et al. 2001)
Can the above problems be solved with a correct picture of the 
obscuration?

\begin{figure}
\centerline{\psfig{file=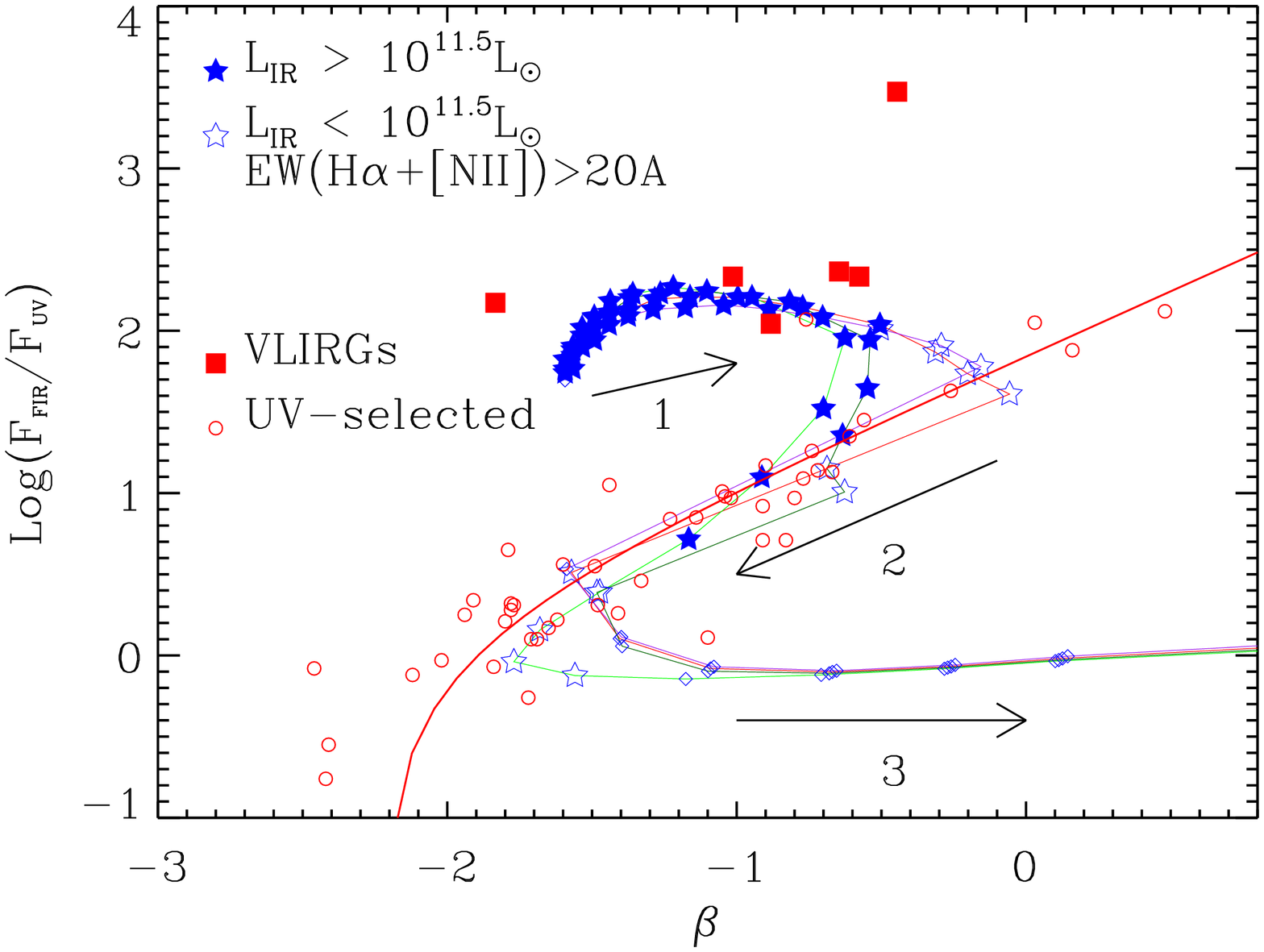,width=6.4truecm,height=5.4truecm,angle=90}
\psfig{file=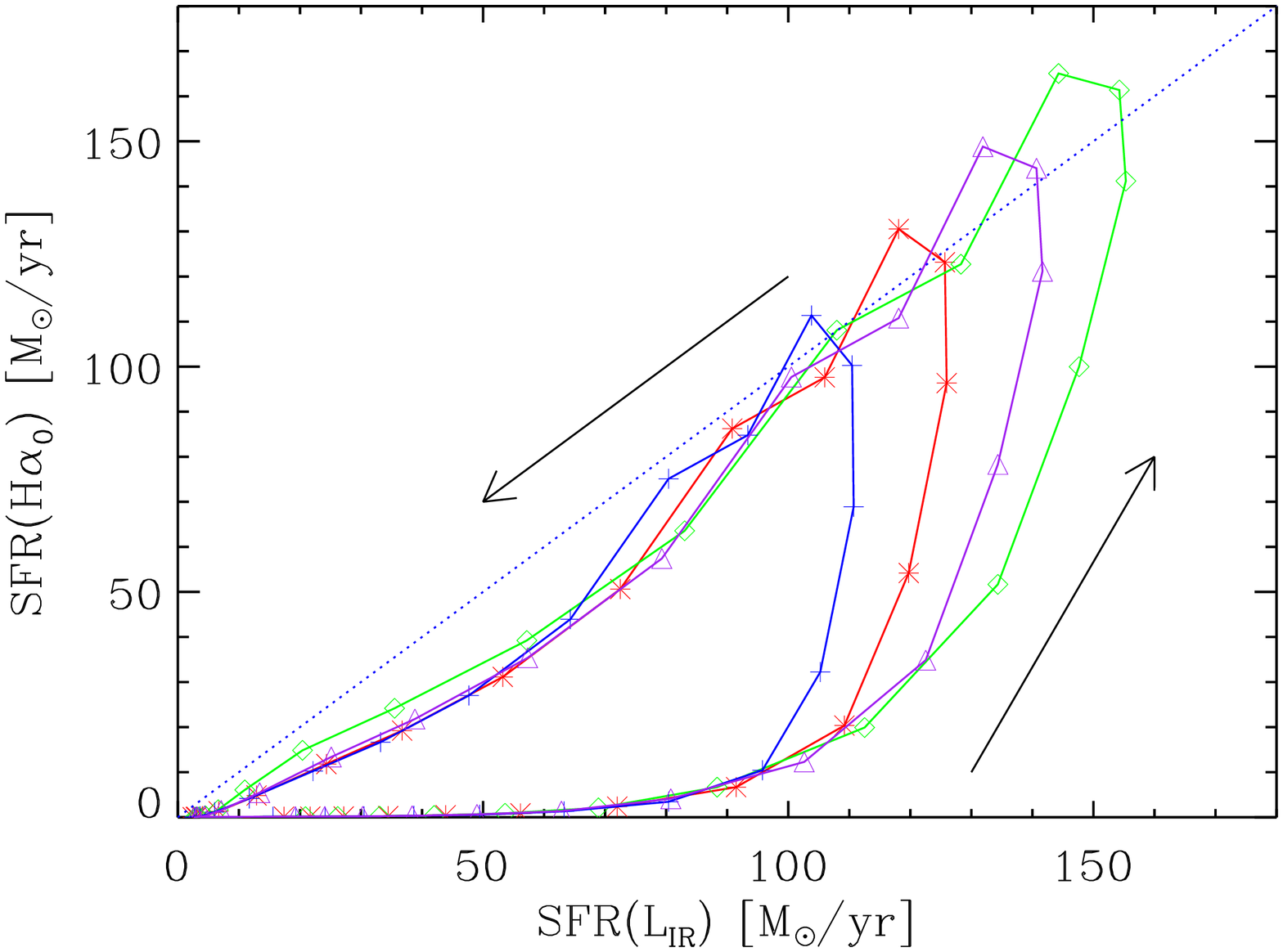,width=5.8truecm,height=5.4truecm,angle=90}}
\caption[]{{\em Left:} The time evolution of the model in
$\beta$-FIR/FUV plane. Observed data are ULIRG
observations (filed squares, Meurer et al. 2000) and UV-selected starburst
(open circles, Meurer et al. 2000). Different paths correspond to 
different fraction of gas converted in stars.  
{\em Right:} SFR derived from H$\alpha$ corrected for
extinction with Balmer decrement versus SFR from FIR.}
\label{risultati}
\end{figure}

We have simulated with our code a starburst galaxy
(barionic mass of $5\cdot 10^{10} M_\odot$) with a
quiescent star formation lasting all its history, plus a final
(analitycal) burst.
We supposed that, as the burst begins, half of total
($1.2\cdot 10^{10} M_\odot$) gas
is under molecular form, and a fraction (from 70\% to
99\%) of it is converted in stars during the burst.
As the molecular gas is consumed during the burst, the MCs
become more and more transparent. This is to mimic the
consumption of the gas and the feedback of SNe.
The initial optical thickness of MC is chosen to
reproduce observed eq. widht of H$\alpha$+[NII] and
H$\beta$ (-62.5 \AA\ and 0.7 \AA) and the ratio $L_{\rm
FIR}/L_{\rm V}$ (88, Poggianti et al., 2001). %\cite{pog01}
The time evolution of the model is shown in figure \ref{risultati}.

% (The picture can be more complex 
%by considering the feedback of SN)

%%%%%%%%%%%%%%%%%%%%%%%%%%%%%%%%%%%%%%%%%%%

%\centerline{\psfig{file=meurer_birx.ps,width=12truecm,angle=90}}
%\centerline{\psfig{file=bianca_sirsha.ps,width=12truecm,angle=90}}

%\begin{figure}
%\centerline{\psfig{file=sfr_old.ps,width=8truecm,height=5truecm,angle=90}\psfig{file=mg_old.ps,width=8truecm,height=5truecm,angle=90}}
%\centerline{\psfig{file=z_old.ps,width=8truecm,height=5truecm,angle=90}\psfig{file=sfr_new_1_u.ps,width=8truecm,height=5truecm,angle=90}}
%\end{figure}

%astro-ph/0011201
%\section{Results}

The temporal evolution in the $\beta$-FIR/FUV plane can
be described in three phases:\\
{ 1)} VLIRG phase: the burst is very
obscured; IR emission comes from the burst while UV comes
from the disk. There is no correlation between UV and IR
because they come from different component of the galaxy.\\ 
{ 2)} UV-bright phase: the burst has almost
consumed the gas and the
MCs become transparent; the model moves towards Meurer's relation.\\
{ 3)} Final phase: the burst is aging and
the model departs from the relation. The galaxy can not be
observationally selected as a starburst any longer.

%During the initial VLIRG phase, the model has the same
%spectral properties of e(a) galaxies, and we confirm the
%nature of obscured starburst.
In inital VLIRG phase,
H$\alpha$ corrected for extinction from
Balmer decrement H$\alpha$/H$\beta$ 
can be a wrong estimate of the ``true'' H$\alpha$ (and
then of the SFR, see fig. \ref{risultati} right). In fact in this phase
H$\beta$ can be dominated by the population outside 
MCs affected by a low extinction; then, the Balmer
decrement is altered because of the selective extinction.
This effect can explain the low $L_{{\rm
H}\alpha_0}/L_{Bol}$ observed in VLIRGs \cite{meu01}.
Concluding, we are able to interpret different UV, optical, IR properties of
obscured galaxies within a unique star formation
selective (in age) extinction scenario. We found that
the relation $\beta$ -- FIR/FUV is not unique in the
obscured phase. For the same reason, the correction of
H$\alpha$ from balmer decrement can be wrong in this phase.

%\centerline{\psfig{file=timeir_1_u_v.ps,width=4truecm,angle=90}}
%\centerline{\psfig{file=timeha_1_u_v.ps,width=4truecm,angle=90}}

%%%%%%%%%%%%%%%%%%%%%%%%%%%%%%%%%%%%%%%%%%

%\centerline{\psfig{file=sirsha_u_v.ps,width=4truecm,angle=90}}
%\centerline{\psfig{file=sirsha0_u_v.ps,width=4truecm,angle=90}}

%\centerline{\psfig{file=mmc_new_u.ps,width=4truecm,angle=90}}
%\centerline{\psfig{file=ebv_new_u.ps,width=4truecm,angle=90}}

%\centerline{\psfig{file=ebvebv_1_u_v.ps,width=4truecm,angle=90}}

%%%%%%%%%%%%%%%%%%%%%%%%%%%%%%%%%%%%%%%%%%

\end{article}

\end{document}